\documentclass[aps,prx,twocolumn,floatfix,superscriptaddress,longbibliography,showpacs,amsmath,amssymb]{revtex4-2}
\usepackage{graphicx}
\usepackage{epsfig}
\usepackage{hyperref}
\usepackage{color,colordvi}
\usepackage{float}

\newcommand{\eps}{\varepsilon} 
\newcommand{\aver}[1]{\langle {#1} \rangle}
\newcommand{\es}[1]{\begin{split}#1\end{split}}
\newcommand{\beq}{\begin{equation}}
\newcommand{\eeq}{\end{equation}}

\newcommand{\lp}{\left(}
\newcommand{\rp}{\right)}
\newcommand{\lsq}{\left[}
\newcommand{\rsq}{\right]}
\newcommand{\lbr}{\left\lbrace}
\newcommand{\rbr}{\right\rbrace}
\newcommand{\da}{\dagger}
\newcommand{\bma}{\begin{pmatrix}}
\newcommand{\ema}{\end{pmatrix}}

\newcommand{\bra}[1]{\langle #1 |}
\newcommand{\ket}[1]{| #1 \rangle}
\newcommand{\mo}{{-1}}
\newcommand{\rw}{\rightarrow}
\newcommand{\oh}{\frac{1}{2}}
\newcommand{\w}{\omega}

\newcommand{\re}{\text{Re}}
\newcommand{\im}{\text{Im}}
\newcommand{\tr}{\text{tr}}
\newcommand{\abs}[1]{ \left\lvert #1	\right\rvert}

\newcommand{\hc}{\text{H.c.}}

\newcommand{\pt}{\partial _t}

\newcommand{\eff}{\text{eff}}

\begin{document}

\title{On the stability of dissipatively-prepared Mott insulators of photons}

\author{Orazio Scarlatella}
\affiliation{T.C.M. Group, Cavendish Laboratory, University of Cambridge, J.J. Thomson Avenue, Cambridge CB3 0HE, UK}
\affiliation{Clarendon Laboratory, University of Oxford, Parks
Road, Oxford OX1 3PU, UK}
\author{Aashish A. Clerk}
\affiliation{Pritzker School of Molecular Engineering, University of Chicago,
5640 South Ellis Avenue, Chicago, Illinois 60637, USA}
\author{Marco Schir\`o}
\affiliation{JEIP, UAR 3573 CNRS, Coll\`{e}ge de France, PSL Research University, 11 Place Marcelin Berthelot, 75321 Paris Cedex 05, France}

\begin{abstract}
Reservoir engineering is a powerful approach for using controlled driven-dissipative dynamics to prepare target quantum states and phases.  
In this work, we study a paradigmatic model that can realize a Mott insulator of photons in its steady-state. We show that, while in some regimes its steady state approximates
a Mott-insulating ground state, this phase can become unstable through a {\it non-equilibrium} transition towards a coherent yet non-classical limit-cycle phase, driven by doublon excitations.  This instability is completely distinct from the ground-state Mott-insulator to superfluid transition.  
This difference has dramatic observable consequences and leads to an intrinsic fragility of the steady-state Mott phase: a fast pump compared to losses is required to sustain the phase, but also determines a small critical hopping. 
We identify unique features of the steady-state Mott phase and its instability, 
that distinguish them from their ground-state counterpart and can be measured in experiments.
\end{abstract}

\maketitle


\section{Introduction}

Dissipation engineering offers a promising avenue for the control of quantum devices and simulators, and is actively being explored as a strategy for stabilizing entangled resource states and phases of matter~\cite{poyatosZoller1996,plenioKnight1999,diehl2008quantum,verstraete2009quantum,Harrington2022engineered}. 
While many physical platforms have been considered, 
superconducting circuit QED systems~\cite{schoelkopfGirvin2008,devoretSchoelkopf2013,blais2021circuit} are particularly ideal for engineering tailored reservoirs.  Reservoir engineering has been widely employed in these systems, including for quantum-state preparation \cite{dassonnevilleHuard2021,
leghtasDevoret2015,
andersenEichler2019,
luSchuster2017a,
kimchi-schwartzSiddiqi2016,
hollandSchoelkopf2015,
lescanne2020exponential} and autonomous quantum error correction \cite{puttermanNoh2022,
sivakDevoret2023,
berdouLeghtas2023,
grimmDevoret2020,
puriBlais2017,
kapitKapit2017,
touzardDevoret2018a}. 
These approaches could be combined with the ability to wire up superconducting circuits into arrays, providing new avenues for  photonic quantum simulators enjoying strong non-linearities and long lifetimes~\cite{houckKoch2012,carusottoSimon2020}.

A recent breakthrough experiment \cite{maSchuster2019} succeeded in dissipatively preparing the first Mott insulator of photons, potentially enabling the realization of correlated quantum fluids of light~\cite{carusotto2013quantum}. 
In this experiment a Bose-Hubbard lattice was realised using an array of transmon qubits, which was then pumped by a reservoir, at a rate faster than lattice losses and that is structured in energy, providing an effective chemical potential for microwave photons~\cite{kapitSimon2014a,hafeziTaylor2015,lebreuillyCarusotto2017,maSimon2017}.
The experiment
demonstrated that the steady state of the dynamics was
a low-entropy incompressible state with integer filling approximating
a ground-state Mott insulator \cite{fisher89boson}.
Despite this achievement, many basic properties of this dissipatively-prepared Mott insulator remain unexplored even at the level of theory.  This includes its spectral properties and the nature of phase transitions out of the Mott state.

In this Letter, we identify a new mechanism by which the steady-state Mott phase can become unstable. While the pumping stops autonomously once the Mott steady-state is reached at small values of hopping, thanks to its finite bandwidth and to the gapped nature of the phase \cite{maSchuster2019}, beyond a critical hopping an instability takes place, characterized by a sudden proliferation of doublon excitations, leading to the onset of a limit-cycle phase, destroying the Mott order. This latter phase is coherent, yet inherits non-classical features from the Mott phase. 

The non-equilibrium transition we unveiled is completely distinct from the ground-state Mott-superfluid one, and has a qualitatively different phase diagram. In particular, we find that a fast pump compared to losses, that is required to sustain the Mott steady-state, translates into a small critical hopping, determining a trade-off between fidelity and stability of this phase. 
Furthermore, we identify unique features of the steady-state Mott phase and its instability, such as amplification gain, a proliferation of doublon excitations and a diverging susceptibility at their energy, that distinguish them from their ground-state counterpart and could be measured in experiments.

Our work differs from previously identified instabilities of steady-state Mott phases. 
In \cite{biellaCiuti2017,caleffiCarusotto2023} a different limit-cycle instability is predicted, driven by hole excitations, forming due to an inefficient pumping scheme. This is similar to the ground-state transition of hard-core bosons which is driven solely by an incommensurate filling \cite{krauthTrivedi1991},\cite{schmidDorneich2002}, rather than by a competition between kinetic term and local repulsion, that instead characterizes the transition we identify here. 
A pumping scheme similar to our manuscript was considered instead in Ref.~\cite{lebreuillyCarusotto2017}, but this reference did not predict the non-equilibrium instability we find here, but rather one similar to the ground-state Mott-superfluid transition. 




We also note that limit-cycle instabilities of driven-dissipative bosons were also discussed elsewhere \cite{scarlatellaSchiro2019,scarlatellaSchiro2021,szymanskaLittlewood2006}, but these works did not consider steady-state Mott insulators.


\begin{figure*}[t]
\centering
\includegraphics[width=1\linewidth]{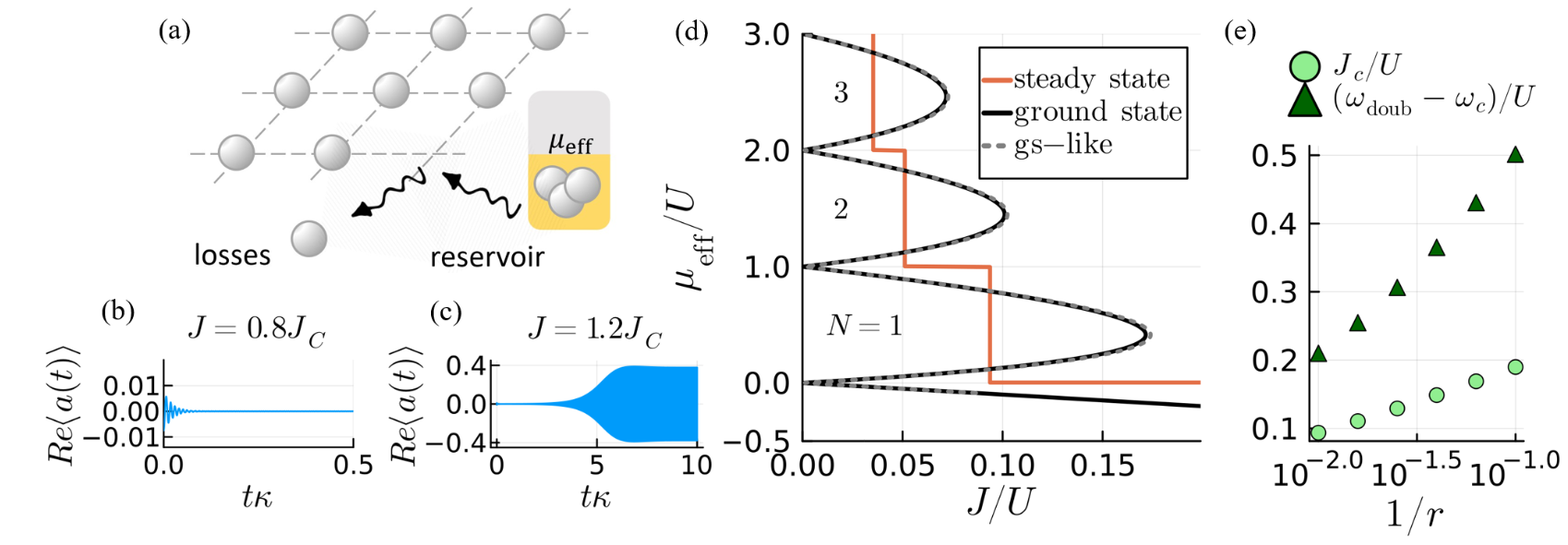}   
\caption{(a) Schematic: a lossy Bose-Hubbard lattice is coupled to a structured reservoir, imposing an effective chemical potential $\mu_{\rm eff}$. 
(b-c) Gutzwiller mean-field (MF) dynamics of the order parameter: for a hopping smaller than a critical value $J<J_c$ (b) it features damped oscillations towards the incoherent ($\aver{a} = 0$) Mott steady-state, while for $J>J_c$ (c) it develops finite-amplitude oscillations corresponding to a coherent limit-cycle phase;
the initial state is a coherent state  $\ket{\alpha}$ with $\alpha=0.01$, and  $J_c/U = 0.094$. 
(d) The steady-state phase diagram in MF: the Mott phase appearing in ``lobes'' with $J<J_c$  with approximately integer filling $N$;
the ground-state Mott-superfluid phase diagram is plotted for comparison, as well as ``ground-state-like'' transition, assuming a static order parameter. 
(e) as the pump/loss ratio $r$ increases, $J_c$ decreases and the critical frequency $\w_c$ approaches the energy $\w_{\rm doub}$ to create a doublon excitation. $\kappa/U = 10^{-3}$ in (a,c) and $\kappa/U = 10^{-5}$ in (d,e). $\mu_{\rm eff}/U = 1/2$ and $r=100$ where relevant.  
}
\label{fig:phaseDiag}
\end{figure*}

\section{Results}
\subsection{The model}
We consider a lossy Bose-Hubbard lattice, incoherently pumped by a structured reservoir, providing a source of incoherent excitations within a finite energy window: this will generate an effective chemical potential for the lattice. The reservoir mimicks the essential features of experimental protocols such as in \cite{maSchuster2019}. 
A minimal model for the system is given by the Lindblad equation
\beq
\partial_t \hat{\rho}=-i[\hat{H}, \hat{\rho}] + \kappa \sum_{j} \mathcal{D} [\hat{a}_{j}] \hat{\rho}  + r \kappa
\sum_{j,\omega} S (\w)  \mathcal{D} [\hat{A}^\da_{j}(\w) ] \hat{\rho}  
\label{eq:mbME}
\eeq
where $\mathcal{D}[\hat{O}] \hat{\rho}=(\hat{O} \hat{\rho} \hat{O}^{\dagger}-\{\hat{O}^{\dagger} \hat{O}, \hat{\rho}\} / 2)$ is a standard Lindblad dissipator.
$\hat{H}$ is the Bose-Hubbard Hamiltonian 
\beq 
\label{eq:hamBh}
\hat{H}= \sum_i  \frac{U}{2}\hat{n}_i(\hat{n}_i-1)  -\frac{J}{z} \sum_{\langle ij\rangle}\left(\hat{a}^{\da}_i \hat{a}_j+\hc \right)
\eeq
where each lattice site hosts a single bosonic mode with annihilation operator $\hat{a}_i$ (and $\hat{n}_i = \hat{a}^\da_i \hat{a}_i$), with an oscillator frequency $\w_0$ that has been gauged away from the Hamiltonian by a rotating frame transformation, $U$ is the on-site interaction, $J$ is the hopping rate between nearest-neighbour sites and $z$ the lattice coordination.

The first dissipator in Eq.~(\ref{eq:mbME}) describes linear, Markovian on-site losses at rate $\kappa$.
The second instead describes the coupling to the structured reservoir, with a pump rate $r \kappa$, such that $r$ is the ratio between pump and loss rates. The jump operators $\hat{A}_{j}^\da(\omega) = \sum_{\varepsilon^{\prime}-\varepsilon=\omega} \hat{\Pi}(\varepsilon^\prime)   \hat{a}_j^\da  \hat{\Pi}\left(\varepsilon\right)$ connect the manifolds of eigenstates of the Hamiltonian differing by one particle, and with energy difference $\varepsilon^{\prime}-\varepsilon=\omega$ (of order $\w_0$ assumed positive and large), where $\hat{\Pi}(\varepsilon)$ is the projector to the manifold with energy $\varepsilon$. $S(\omega)$ is the spectral function of the reservoir \cite{breuerPetruccione2007} for which we use the simple form
\beq
\label{eq:lesserBox}
S(\omega) = \theta(\mu_{\rm eff}-{\w}) 
\eeq 
with $\mu_{\rm eff}$ representing the maximum energy of reservoir excitations (in the rotating frame of the Hamiltonian \eqref{eq:hamBh}). 


The master equation \eqref{eq:mbME} can lead to a steady-state that approximates a ground-state Mott insulator. 
In fact, the structured reservoir with spectral function \eqref{eq:lesserBox} 
approximates the detailed balance relation
characterizing a system in contact with an equilibrium reservoir with chemical potential  $\mu_{\rm eff}$ \cite{lebreuillyCarusotto2017} (see also Appendix \ref{app:equilibration-singSite-gs}): it therefore imposes an effective chemical potential.
The pump/loss ratio $r$ plays the role of an inverse temperature, meaning that the ground-state corresponds to $r \rightarrow \infty$. In practice, the larger $r$, the better the fidelity of the steady-state with the Mott ground-state; this was also as numerically verified by \cite{lebreuillyCarusotto2017}.

In the limit of disconnected sites $J=0$ one can also calculate the steady-state analytically (as shown in Appendix \ref{app:equilibration-singSite-gs}) and check that it corresponds, for large $r$, to the ground state of the Hamiltonian with an equilibrium chemical potential $\mu = \mu_{\rm eff}$, namely a Fock state $\ket{N}$ with filling set by the chemical potential as 
\beq
\label{eq:neqPopCond}
N - 1 < \frac{\mu_{\rm eff}}{U} < N
\eeq
\subsection{Instability of the steady-state Mott phase}
Although for weak $J$ the steady state well approximates a ground-state Mott insulator, the instability of this phase at a critical hopping turns out to be {\it qualitatively} different from the equilibrium case. 

A simple way to capture this instability is to solve the dynamics with a time-dependent Gutzwiller mean-field ansatz $\rho(t)= \prod_i \rho_i(t)$, that neglects both quantum and classical correlations between sites while capturing the local physics. 
This corresponds to solving a single-site problem with an effective Hamiltonian $H_{\rm MF} = U n(n-1)/2  +\phi^{\dagger}(t) a+\phi(t) a^{\dagger}$, where $\phi(t)=-J \langle a(t)\rangle$, where we dropped the site index assuming also a homogeneous-in-space state and in the thermodynamic limit of infinitely-many sites. 
Fig. \ref{fig:phaseDiag}(b) shows that below a critical hopping $J<J_c$ the order parameter, that is the average bosonic field $\aver{a(t)}$,  features damped oscillations and eventually vanishes in the steady-state Mott phase. 
At a critical hopping $J_c$ this develops limit-cycle oscillations at a critical frequency $\w_c$, $\aver{a(t \rw \infty)} = |a| e^{i \omega_c t + i \phi} $, that acquire a stationary and finite amplitude $|a|$ for $J>J_c$.  
This is shown in panel (c); note that for $J>J_c$ the frequency is renormalized with respect to that for $J=J_c$ \cite{scarlatellaSchiro2019}.
We anticipate that this second-order dissipative phase transition is a genuine non-equilibrium instability, distinct from the ground-state Mott-superfluid transition: here the onset of limit-cycles corresponds to a spontaneously broken time-translation symmetry, along with the U(1) symmetry of \eqref{eq:mbME} for $\hat{a}_i \rw \hat{a}_i e^{i\theta}$, something that cannot happen for ground-state transitions \cite{nozieresNozieres2013,brunoBruno2013b,
watanabeOshikawa2015a}. 


The phase boundary can be found by the condition that the lattice susceptibility to an applied weak coherent field diverges at the instability. 
We compute this quantity using a strong-coupling Keldysh field theory approach (detailed in Appendix~\ref{app:RPA}). 
This in principle allows to describe the excitations of the Mott phase, but not of the superfluid phase \cite{senguptaDupuis2005b}, and to formulate a critical theory of the Mott-superfluid transition \cite{scarlatellaSchiro2019,fisher89boson}. It gives the same critical point as Gutzwiller mean-field (as shown in Appendix \ref{app:gutz}), that follows from
\begin{align}
\label{eq:neqCritFreq}
 0 &= - \frac{1}{\pi} \im G_0^R(\omega_c) \\
 \label{eq:neqCritHop}
1/{J_{c}} &= - \re G_0^R(\omega_c)
\end{align}
where $G_0^R(\omega) = -i \int_{0}^\infty dt e^{i \w t } \aver{[ \hat{a}(t),\hat{a}^\da(0) ]}_0 $ is the Fourier transform of the local steady-state susceptibility, evaluated at $J=0$ corresponding to disconnected sites. Note that here homogeneity in space is a result of calculations, rather than an assumption.

Eq.~(\ref{eq:neqCritFreq}) determines the Fourier mode that becomes unstable, identified by the critical frequency $\w_c$. It represents a zero net-damping condition: on any given lattice site, the rest of the lattice (viewed as a bath) does not produce any net gain or loss. 
As we discuss later, 
thermal equilibrium in the grand-canonical ensemble would require $\w_c =\mu_{\rm eff}$, corresponding to an unbroken time-translational symmetry. 
A violation of this condition can only occur out of equilibrium. 
Eq.~(\ref{eq:neqCritHop}) instead determines the critical hopping $J_c$ from the real part of the susceptibility evaluated at $\w_c$.


The steady-state phase boundary is plotted in Fig.~\ref{fig:phaseDiag}(d), showing that the regimes of stable Mott phases form ``lobes'' in the hopping-chemical potential plane.  Each distinct lobe corresponds to a chemical potential range 
\eqref{eq:neqPopCond} where there is an approximate integer filling of the lattice.  One immediately sees that their shape is dramatically altered in the steady-state case, as compared to the standard ground-state Mott-superfluid transition (also plotted).  We discuss later how those difference arise, highlighting first the main consequences. 
An expected result is that dissipation suppresses phase coherence in parts of the phase diagram, as we see small regions that would be superfluid in the ground state turned into Mott insulators in the steady state.
Conversely, it is remarkable that for many values of $\mu_{\rm eff}$ in the plot the Mott steady-state is {\it unstable} at a smaller critical hopping $J_c$.

An important results is that the critical hopping $J_c$ strongly depends on the pump/loss ratio $r$, and decreases when increasing $r$, as Fig.~\ref{fig:phaseDiag}(e) shows. 
This implies a trade off between the fidelity of the steady-state to a ground-state Mott insulator, which requires a fast pump compared to losses (i.e. a large $r$) to sustain an integer filling, as discussed earlier in the manuscript and verified numerically in \cite{lebreuillyCarusotto2017}, and the stability of this phase at finite hopping. 

Finally, Fig.~\ref{fig:phaseDiag}(e) also shows that the critical frequency \eqref{eq:neqCritFreq} 
approaches for large $r$ the energy (using $\hbar = 1$ units) to create a doublon excitation, namely to add one particle to the steady-state, $\w_c \sim \w_{\rm doub}$, that in a single-site picture is
\beq
\label{eq:doubEn}
\w_{\rm doub}  =  U  N  
\eeq 
where $N$ is the filling of the Mott state.

\begin{figure}[t]
\centering
\includegraphics[width=1\linewidth]{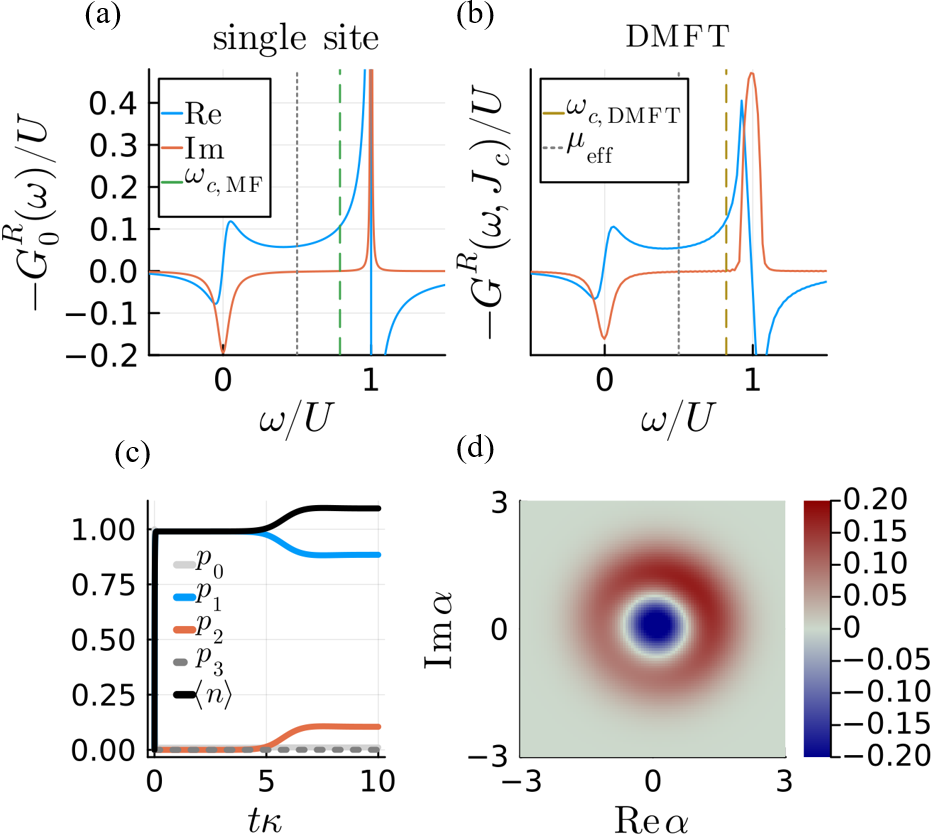} 
\caption{
 Local susceptibility for $\mu_{\rm eff}/U = 1/2$ for (a) the single-site problem and (b) in DMFT, on a Bethe lattice with coordination $z=20$ and at $J_{c,\rm dmft}/U = 0.086$.
The critical frequency $\w_c$ is marked, both in MF and DMFT and for the ground-state transition (dotted-gray), identifying a region of negative density of states (NDoS) in between.
(c) Dynamics of Fock states populations $p_n(t) = \bra{n} \rho(t) \ket{n}$ and density $\aver{n(t)}$ in the limit-cycle phase in MF: at the onset of limit cycles (for $t\kappa \approx 6$) the doublons $(p_2)$ and the density suddenly increase.
(d) Wigner function of the limit-cycle state at $t\kappa=8.3$ in MF: lack of rotational symmetry signals phase coherence. The negative peak signals a non-classical state. For all panels
$\mu_{\rm eff}/U = 1/2$,  $\kappa/U = 10^{-3}$ and $r=100$ as in Fig.~\ref{fig:phaseDiag}(a,c).
}
\label{fig:dmft}
\end{figure}

To give more insights on the instability, in Fig.~\ref{fig:dmft}a we plot the susceptibility $G_0^R(\omega)$ that controls it via \eqref{eq:neqCritFreq},\eqref{eq:neqCritHop}. 
Its imaginary part shows two peaks, from left to right respectively at the energy of hole and doublon \eqref{eq:doubEn} excitations, corresponding to Mott-Hubbard bands separated by a gap $U$. 
We see that the critical frequency (dashed line) is indeed close to the bottom of the upper Hubbard band, around the doublon energy.
We can also understand from this figure the related behaviour of $\w_c$ and $J_c = -1/ReG^R_0(\w_c)$  as functions of $r$ already discussed:  the real part becomes larger if evaluated at a critical frequency closer to $\w_{\rm doub}$ \eqref{eq:doubEn}; in Appendix~\ref{app:doublonResonance}, we trace this back to the Kramer-Kronig relations of susceptibilities. 
Physically, $\w_c \sim \w_{\rm doub}$ reflects into the instability being characterized by a sudden generation of doublons in the dynamics for $J>J_c$, accompanied by an increase in the density: this is shown in  Fig.~\ref{fig:dmft}(c).
This also corresponds to the onset of phase coherence, which is confirmed by Fig.~\ref{fig:dmft}(d) showing the Wigner function of the limit cycle state: it is not circularly symmetric.  It nonetheless possesses a strong non-classical character, signalled by the negative peak of the Wigner function near the origin, reminiscent of the Mott phase.

Another aspect of the local susceptibilities is important to remark. Their imaginary part (with a minus sign), is a local probe of the density of states (DoS). 
Our instability requires the system to exhibit a negative DoS (NDoS) at frequencies $\w > \mu_{\rm eff}$,
something that is directly connected with the ability to generate amplification gain and that cannot occur in equilibrium conditions~\cite{scarlatellaSchiro2019a,scarlatellaSchiro2021}. This happens for $\mu_{\rm eff} < \w < \w_c$  in our case as indicated in Fig.~\ref{fig:dmft}a,b by the dashed and dotted lines (though its amplitude is not appreciable with the scale used).
The local NDoS directly reflects in a diverging lattice susceptibility at zero momentum and at the critical point 
\eqref{eq:neqCritFreq}\eqref{eq:neqCritHop}.  
The NDoS, the diverging susceptibility at $\w_c \sim \w_{\rm doub}$ and the 
increase in time of doublons and density (Fig. \ref{fig:dmft}c)
are unique properties of the steady-state Mott phase, that distinguish it from a ground-state Mott insulator. These can be probed in experiments by measurements of transmission/reflection and density 
 (see e.g. \cite{magazzuGrifoni2018,maSchuster2019}) .

\subsection{Comparison with ground-state Mott-superfluid transition}
For comparison, note that the ground-state Mott-superfluid transition is also described by the Hamiltonian \eqref{eq:hamBh}, up to a rotating frame transformation equivalent to a standard grand-canonical Hamiltonian with equilibrium chemical potential $\mu_{\rm eff}$.
Also, the critical point Eqs.~ \eqref{eq:neqCritFreq}\eqref{eq:neqCritHop} hold as well for the ground-state transition: one simply replaces the steady-state susceptibility with the analogous ground-state quantity.

The main difference is that in the ground-state case, only a static order parameter can form, since time-translation symmetry cannot be broken spontaneously  \cite{nozieresNozieres2013,brunoBruno2013b,watanabeOshikawa2015a}.  This constraints the critical frequency to take the equilibrium value $\w_c=\mu_{\rm eff}$ (in the rotating frame of \eqref{eq:hamBh}). 
Accordingly, Eq. \eqref{eq:neqCritFreq}  for the ground-state is guaranteed to be satisfied at this frequency \cite{scarlatellaSchiro2019a}.
Note that the $\mu_{\rm eff}$-dependence of $\w_c$ gives rise to the the typical ``round'' Mott lobes. Ground-state calculations are reported in Appendix~\ref{app:equilibration-singSite-gs}. 

In a non-equilibrium scenario instead, $\w_c$ is not a priori known: in our case, this must be found from \eqref{eq:neqCritFreq}. 
This predicts a very different value from equilibrium: in particular $\w_c$ strongly depends on the pump/loss ratio $r$, approaching the doublon frequency \eqref{eq:doubEn} for large $r$ as previously shown in Fig 1e; also it does not depend on $\mu_{\rm eff}$ within one lobe, and thus $J_c$ also doesn't, giving rise to the ``flat'' lobes. 

Eventually, the large difference of the steady-state phase diagram compared to the ground-state one (Fig.  \ref{fig:phaseDiag}d) is due to the critical hopping equation \eqref{eq:neqCritHop} being evaluated at a very different critical frequency from equilibrium (the dashed rather then dotted line in Fig. \ref{fig:dmft}a,b). 
Indeed, if the equilibrium frequency is assumed, this yields a phase diagram that only differs perturbatively in the dissipation strength from the ground-state one: the dotted line in Fig.  \ref{fig:phaseDiag}d.

A similar result if found in \cite{lebreuillyCarusotto2017}, that making a similar assumption of a static transition for a similar problem, finds a ground-state-like phase diagram. Our theory is therefore consistent with \cite{lebreuillyCarusotto2017}. 
Also, we expect our instability, predicted in the thermodynamic limit, to be absent in the steady-state of a finite-size system like in \cite{lebreuillyCarusotto2017}, as the limit-cycle would likely become a long-lived metastable state in this case, with a lifetime that only diverges in the thermodynamic limit. 



We remark that the $\mu_{\rm eff}$-independent lobes are actually an artifact of our theoretical approaches, relying on the solution of a single-site ($J=0$) Lindblad equation: this can only capture a step-wise dependence on $\mu_{\rm eff}$ (more details in Appendix \ref{app:Mott-indepLobes}). 
While we expect some $\mu_{\rm eff}-$dependence beyond those approaches, the steady-state phase diagram will still retain its characteristic dependence on pump/loss ratio $r$, remaining distinct from the ground-state one. 
Preliminary evidence also suggests that at very large $r$, but still weak overall dissipation, a Lindblad equation might not be accurate \eqref{eq:mbME}.  
Finally, correlations due to dimensionality beyond our approaches, especially in 1D and 2D, might also introduce corrections to the predicted phase diagram.

\subsection{Benchmark of main results}
While we considered a square bath spectral function \eqref{eq:lesserBox}
and a Lindblad master equation \eqref{eq:mbME}, our results do not depend on these choices: in Appendices \ref{app:redfield} and \ref{app:lorentzianRes} similar results are found with a different spectral function and with a Redfield equation.

Here instead we confirm our results going beyond the previous Gutzwiller and Keldysh approaches to the lattice problem using dynamical mean-field theory (DMFT) \cite{scarlatellaSchiro2021,aokiWerner2014,georgesKotliar1992}, with an impurity solver based on the non-crossing approximation ~\cite{schiro2019quantum,scarlatella2021self}.
For bosons \cite{andersWerner2011,byczukVollhardt2008,strandWerner2015,strandWerner2015a}, DMFT captures non-perturbatively the leading $1/z$ corrections to Gutzwiller, where $z$ is the lattice connectivity. In particular, we predicts the local susceptibility and critical point beyond previous approaches \cite{scarlatellaSchiro2021}. 
The local susceptibility at finite hopping  $G^R(\omega,J) = -i \int_{0}^\infty dt e^{i \w t } \aver{[ \hat{a}(t),\hat{a}^\da(0) ]} $ is shown in Fig.~\ref{fig:dmft}b, where one notices the formation of bands replacing the single-site resonances of panel (a).  
We assumed for simplicity a Bethe lattice \cite{georgesRozenberg1996,strandWerner2015} and a homogeneous phase, therefore dropping the site index. 
Similar critical point equations to \eqref{eq:neqCritFreq},\eqref{eq:neqCritHop} exist in DMFT (reported in Appendix \ref{app:dmft}), involving now the $J$-dependent local susceptibility: solving those we confirm our instability, with a critical frequency $\omega_c$ at the bottom of the doublon band (marked in Fig.~\ref{fig:dmft}). 


\section{Conclusions}
We identified a new instability of a dissipatively-prepared Mott insulator, driven by doublon excitations. This is physically different from the ground-state Mott-superfluid transition and is characterized by a dependence on the pump strength that reveals an intrinsic trade-off between the fidelity of the steady-state with a Mott state and its stability to finite hopping. 
We connected the steady-state Mott phase and its instability to peculiar features of the dynamics and susceptibilities that can be measured in experiments.
Our results are relevant to the dissipative preparation of gapped phases of matter that can be achieved with similar protocols \cite{maSchuster2019}, beyond the specific case of a Mott phase studied here, see for example Ref.~\cite{mi2023stable}.


\section*{Acknowledgements}

We acknowledge discussions with Jonathan Simon, Andrei Vrajitoarea, Gabrielle Roberts and Meg Panetta at University of Chicago and Stanford University.
This work was supported by the Air Force Office of Scientific Research MURI program under Grant No. FA9550-19-1-0399, and the Simons Foundation through a Simons Investigator award (Grant No. 669487).
It was also supported by the Engineering and Physical Sciences Research Council [grant number EP/W005484] and by the
European Research Council under the European Union’s
Seventh Framework Programme (FP7/2007-2013)/ERC
Grant Agreement No. 319286 Q-MAC. 
MS acknowledges support from the ANR grant ``NonEQuMat'' (ANR-19-CE47-0001).
This work was supported by the University of Chicago through a FACCTS grant (``France and Chicago Collaborating in The Science'').
For the purpose of open access, the authors has applied a creative commons attribution (CC BY) licence to any author accepted manuscript version arising.
The data (code) to reproduce the results of the manuscript will provided by the author under request.




\appendix

\section{Equilibration, single-site problem and comparison with the ground-state case}
\label{app:equilibration-singSite-gs}

The master equation \eqref{eq:mbME}, in the regime considered throughout the paper of large pump/loss ratio $r \gg 1$ but weakly coupled environment $\kappa, r \kappa \ll U,J$, and for $U \ll J$, approximates the dynamics of equilibration with a
bath at chemical potential $\mu=\mu_{\rm eff}$ and zero temperature $T=0$: as a consequence, its steady-state is expected to approximate the ground state of a grand-canonical Hamiltonian.
The equilibrium dynamics is in fact characterized by the detailed balance relation, stating that given two eigenstates of the Hamiltonian $\ket{\psi}$ and $\ket{\phi}$, the ratio of the transition rates for going from one state to the other equals the ratio of their equilbrium probabilities, their Boltzmann weights, at temperature $T$ and chemical potential $\mu$: $ {\mathcal{T}^{\rm eq}_{\psi \rw \phi}}/{\mathcal{T}^{\rm eq}_{\phi \rw \psi}} = e^{ (\mu - \eps_\phi  + \eps_\psi)/T }$.
The master equation \eqref{eq:mbME} defines the following transition rates between eigenstates differing by 1 particle: if $\ket{\psi}$ has 1 particle less than $\ket{\phi}$, then $ {\mathcal{T}_{\psi \rw \phi}}/{\mathcal{T}_{\phi \rw \psi}} = r \theta( \mu_{\rm eff} -\eps_\phi +\eps_\psi ) $, that for $r \gg 1$ approximates the detailed balance relation at zero temperature $T\rw 0$ and for $\mu = \mu_{\rm eff}$. A similar discussion is reported in \cite{lebreuillyCarusotto2017}. 
We remark that this partial detailed balance relation does not guarantee an equilibrium steady state, as Eq. \eqref{eq:mbME} does not guarantee thermalization within each fixed particle-number subspace.  Further, in the presence of spectral degeneracies, the master equation will couple populations and coherences.  Nonetheless, the expectation of an equilibrium state is expected to hold if one is deep in the Mott phase, as it becomes rigorous at zero hopping. 


For a single-site problem ($J=0$) the steady-state can be calculated analytically and shown to correspond, for large $r$, to the Bose-Hubbard-site ground state with chemical potential $\mu = \mu_{\rm eff}$.
Note that a collection of independent sites is also representative of a Mott insulating phase in the Gutzwiller approximation. 
For the single-site problem, the jump operators entering \eqref{eq:mbME} become simply $ A^\da(E_{n+1} - E_{n}  ) =  \aver{ n+1  | a^\da| n} \ket{n+1} \bra{n} $ (omitting the site index), describing transitions between two Fock states, the single-site eigenstates, differing by one boson with energy difference $E_{n+1} - E_{n} =  U n $. The master equation \eqref{eq:mbME} reduces to a simple rate equation for Fock states populations. Therefore the detailed balance argument discussed above becomes rigorous and the steady-state for large $r$ must correspond to the single-site ground state. 
The single-site master equation takes the form: 
\beq
\partial_t \hat{\rho}=-i[\hat{H}, \hat{\rho}] + \kappa  \mathcal{D} [\hat{a}] \hat{\rho}  + r \kappa
\sum_{n=0}^{\infty} S(nU)  \mathcal{D} [\sqrt{n+1}\ket{n+1}\bra{n}] \hat{\rho}  
\label{eq:singSiteME}
\eeq
where the bath spectral function is still given by \eqref{eq:lesserBox}, $S(\omega) = \theta(\mu_{\rm eff} - \omega)$,  and is evaluated at the single-site energy differences. 

In the steady-state, one finds that only eigenstates with $E_n -E_{n-1} = n U < \mu_{\rm eff}$ are populated and that the populations are given by $p_n = r^n  ({1-r})/({1-r^{N+1}}) \theta(N-n) $ where $N$ is the last populated Fock state satisfying \eqref{eq:neqPopCond}.
For $r\gg 1$ the steady-state approaches the pure state $\ket{N}$, as $p_n \approx \delta_{n,N}$, with $N$ obeying \eqref{eq:neqPopCond}. 

This corresponds to the ground-state of a Bose-Hubbard site with an equilibrium chemical potential $\mu = \mu_{\rm eff}$, i.e. $ \hat{H}_0  = - \mu \hat{n} +  \,  U\hat{n} (\hat{n}-1)/2 $ \cite{fisher89boson,sachdevSachdev2007}. 
The ground-state single-site susceptibility can also be easily computed through a spectral decomposition and reads
\beq
\label{eq:gsRet}
G_{0,\rm gs}^R (\w)  =  \frac{N +1}{ {\w }- \w_{\rm doub}^{\rm gs} + i \eta } - \frac{N }{{\w }- \w_{\rm hol}^{\rm gs} + i \eta }
\eeq
where $\eta$ is a positive infinitesimal.
Like its steady-state counterpart plotted in Fig. \ref{fig:dmft}, it has two peaks at the energies of doublons $\w_{\rm doub}^{\rm gs}$ and holon $\w_{\rm hol}^{\rm gs}$ excitations (corresponding to adding and removing a particle from the ground-state), which are given by 
\begin{align}
\label{eq:gsExc}
\pm \w_{\rm doub(hol)}^{\rm gs} &= \pm (E_{N \pm 1} - E_{N } ) = - {\mu } + U\lp N  - \oh \pm \oh \rp 
\end{align}
Going to a rotating frame in which the chemical potential is removed from the Hamiltonian like in the main-text Hamiltonian \eqref{eq:hamBh} these energies are shifted by $\mu$, and coincide with the steady-state expression \eqref{eq:doubEn} reported in the main text.
%

The ground-state Gutzwiller mean-field phase diagram is given by the main text critical point equations \eqref{eq:neqCritHop}, \eqref{eq:neqCritFreq}, where the steady-state susceptibility is  replaced with the ground-state one \eqref{eq:gsRet}. 
Since the ground-state transition is static, those equations must be evaluated at $\w_{c} = 0$ with the susceptibility \eqref{eq:gsRet} (or equivalently at $\w_c=\mu$ in the rotating frame of \eqref{eq:hamBh}).
Note that \eqref{eq:neqCritFreq} is always satisfied at such a frequency for equilibrium states such as ground-states (see e.g. \cite{scarlatellaSchiro2019a}). The critical hopping is then given by Eq. \eqref{eq:critHopDmft}
yielding the well known Mott lobes in the $\mu-J$ plane, plotted in Fig. \ref{fig:phaseDiag} with $\mu =\mu_{\rm eff}$ for comparison with the steady-state phase boundary.

\section{Details on the Mott phase instability}
\label{app:detailsOnInstability}

Here we discuss how to obtain the critical point equations \eqref{eq:neqCritFreq}\eqref{eq:neqCritHop} for the phase transition out of the Mott phase. 
We first discuss how to obtain it from the Gutzwiller dynamics and then using the strong-coupling RPA Keldysh field theory. Finally, we discuss how these equations are modified within Dynamical Mean-Field Theory.

\subsection{Time-dependent Gutzwiller}
\label{app:gutz}

Within the Gutzwiller ansaz made in the main text, the linear response to a small symmetry-breaking field $\phi(t)$ reads
\begin{equation}
\begin{split}
\aver{a(t)} = &\int_{\infty}^{-\infty} d\tau G_0^R(t-\tau)\phi(\tau) = \\-J
&\int_{\infty}^{-\infty} d\tau G_0^R(t-\tau)
\aver{a(\tau)}
\end{split}
\end{equation}

where in the last step we have used the Gutzwiller self-consistency condition  $\phi(t) = - J\aver{a(t)}$. Translating the above condition in frequency domain we obtain $ \aver{a(\w)} = -J G_0^R(\omega )\aver{a(\w)}$.
For a given $\w$ such that $a(\w) \neq 0$, this equation gives the critical point equations \eqref{eq:neqCritFreq}, \eqref{eq:neqCritHop}: 
\begin{equation}
1/J + G_0^R(\omega ) = 0
\end{equation}
The smallest $J$ and the corresponding $\w$ satisfying this condition define the critical point $(J_c,\w_c)$. 

The same condition can be recovered in a strong-coupling Keldysh field theory \cite{senguptaDupuis2005b}, as we show in the following. 


\subsection{Strong-coupling RPA in the Keldysh path integral}
\label{app:RPA}

We make a strong-coupling RPA (random phase approximation) \cite{scarlatellaSchiro2019,senguptaDupuis2005b}, formulating the problem in the language of Keldysh field theory. The Keldysh action, in terms of the coherent fields $a_i,\bar{a}_i$ reads
\begin{align}
\label{eq:hbDrivenDissAction}
\es{
S &= \int_\mathcal{C} dt \lp \sum_i \bar{a}_i  i \pt a_i - H \rp + \sum_i \lp S_{l,i} + S_{\mu_{\rm eff},i} \rp 
} 
\end{align}
where $\int_\mathcal{C}$ is an integral on the Keldysh contour, $H$ is the expectation value of the Hamiltonian on coherent states 
\beq
H = \sum_i\left( \omega_0 \bar{a}_i a_i +\frac{U}{2} \bar{a}_i \bar{a}_i a_i a_i \right) -  \sum_{\aver{ij}} \frac{J}{z} \, \lp \bar{a}_i  a_j + \hc \rp 
\eeq
and $S_{l,i}$ describes the Markovian losses, corresponding to the loss dissipator 
\beq
\label{eq:1pLossKel}
\es{S_{l,i} &= -i \kappa \int_{-\infty}^{\infty} dt \,  \lp \bar{a}_{i-} a_{i+} -\oh  \bar{a}_{i+} a_{i+} -\oh  \bar{a}_{i-} a_{i-} \rp 
}
\eeq
Finally, $S_{\mu_{\rm eff},i}$ describes the coupling to the structured reservoir. For a bath of non-interacting bosons this can be integrated out explicitly, yielding 
\beq
\label{eq:nmBathKel}
 S_{\mu_{\rm eff},i} =-i  \int_\mathcal{C} dt \int_\mathcal{C} dt' r \kappa \sum_i \bar{a}_i(t) C(t-t') a_i(t')
\eeq
where $C(t-t')$ is the bath correlation function, 
with Keldysh indices implicit in the time variables. 
The fourier transform of the lesser component of this function is $S(\w)$ defined in the main text \eqref{eq:lesserBox}.

We then do a Hubbard-Stratonovich transformation on the hopping term, by introducing the auxiliary bosonic fields $\psi_i$ by a Gaussian integral 
\begin{widetext}
\beq
\exp \lp { {-} i \int_\mathcal{C} dt \sum_{ij} \bar{a}_i J_{ij} a_j} \rp =\frac{1}{\mathcal{N}} \int \prod_{i} \bold{D} \lsq \bar{\psi}_i {\psi}_i \rsq \exp{\lbr i \int_\mathcal{C} dt  \lsq \sum_{ij} \bar{\psi}_i J_{ij}^\mo \psi_j  + \sum_i \lp  \bar{\psi}_i a_i + \psi_i \bar{a}_i \rp \rsq \rbr} 
\eeq

%
%
where $J_{ij}^\mo$ is the inverse hopping matrix and
$\mathcal{N}$ is a normalization coming from Gaussian integration.
By plugging this in the action \eqref{eq:hbDrivenDissAction}, we can formally integrate on the $a_i,\bar{a}_i$ fields 
\beq
\es{
Z &= \int \prod_{i} \bold{D}[  \bar{a}_i ,  a_i ] e^{i S[ \lbr \bar{a}_i , a_i \rbr ]  } = \\
&= \frac{1}{\mathcal{N}} \int \prod_{i} \bold{D}[ \bar{\psi}_i, {\psi}_i ] e^{ i \int_\mathcal{C} dt  \sum_{ij} \bar{\psi}_i J_{ij}^\mo \psi_j } \int \prod_{i} \bold{D}[  \bar{a}_i ,  a_i ] e^{i S_{loc} }  e^{ i \int_\mathcal{C} \sum_i \lp  \bar{\psi}_i a_i + \psi_i \bar{a}_i \rp  } \\ 
&= \frac{1}{\mathcal{N}} \int \prod_{i} \bold{D}[ \bar{\psi}_i, {\psi}_i ] e^{i S_{\rm eff}  [ \lbr \bar{\psi}_i , \psi_i \rbr ] }
}
\eeq
\end{widetext}
obtaining the effective action for the fields $\psi_i, \bar{\psi}_i$ alone
\beq \label{eqn:Seff}
\mathcal{S}_{\eff}= \int_\mathcal{C} dt \lp \sum_{ij}\bar{\psi}_i J_{ij}^{-1}\psi_j+\sum_i\Gamma[\bar{\psi} _i,\psi_i] \rp
\eeq
where the second term represents the generating functional of the bosonic Green functions of isolated sites, $\Gamma[\bar{\psi}_i,\psi_i]= - i \log\langle T_C e^{i\int_\mathcal{C} dt\left(\bar{\psi}_i a_i+a^\da_i \psi_i\right)}\rangle_{0}$.

We stress that the latter average is taken on the single-site problem, therefore the many-body problem has been formally reduced to calculating the single-site problem cumulants. 
We remark that only at this stage we use the description of the reservoir with a Lindblad equation \eqref{eq:mbME}, that therefore only depends on the spectrum of the single-site problem and can be evaluated in practice. 

To obtain the strong-coupling RPA, we then truncate the effective action at the Gaussian level obtaining 
\beq \label{eqn:Seff}
S_{\eff}= \int_\mathcal{C} dt  \int_\mathcal{C} dt' \sum_{ij}   \bar{\psi}_i  \lp J_{ij}^{-1}+G_0(t-t') \rp \psi_j
\eeq 
where $G_0 (t-t') = - i \aver{ T_\mathcal{C}  a_i(t) a_i^\da(t') } $ is the contour-ordered Green function of the single-site problem.
A second-order phase transition is then signalled by a vanishing retarded component of the effective action, corresponding to a diverging susceptibility at the critical point, which in frequency and momentum space reads 
$ 0=1/J_q-G^R_{0}(\omega)$ with $J_q$ the lattice dispersion.
For a hyper-cubic lattice $J_q=-2{J}/{z}\sum_{\alpha=1}^d\cos q_{\alpha}$ and the first unstable mode is the $q=0$ mode (assuming $\re {G^R_{loc}(\w_c)}<0$), leading to the critical point equations \eqref{eq:neqCritFreq} \eqref{eq:neqCritHop}  of the main text: 
\begin{align}
 0 &= \im G_0^R(\omega_c) \\
\frac{1}{ J_c} &= - \re G_0^R(\omega_c) 
\end{align}

The single-site susceptibility $G_0^R(\omega)$ is finally obtained numerically from the single-site Lindblad equation \eqref{eq:singSiteME}. \\

\subsection{Absence of Mott-lobes dependence on effective chemical potential}
\label{app:Mott-indepLobes}
It is important to note that in the  single-site Lindblad equation \eqref{eq:singSiteME} the reservoir spectral function \eqref{eq:lesserBox}, $S(\omega) = \theta(\mu_{\rm eff} - \omega)$, is evaluated at the transition energies of the single-site Hamiltonian $E_{n+1} - E_{n} = nU$, that don't depend on $\mu_{\rm eff}$ (or on $J$). Therefore the single-site problem simply depends in a step-wise manner on $\mu_{\rm eff}$, by steps of $U$. This reflects the blockade effect of the non-linearity $U$, that is the incompressibility of the Mott state, already at single-site level. Though, note that this also implies that, not only single-site populations, but any quantity computed from \eqref{eq:singSiteME} depends in the same step-wise manner on $\mu_{\rm eff}$. 
Further, since all the many-body techniques we use (including DMFT) eventually reduce the lattice problem to the single-site problem \eqref{eq:singSiteME}, also any lattice quantity eventually depends on $\mu_{\rm eff}$ in the same step-wise way. This includes the critical hopping $J_c$, resulting in the flat lobes.

The perfectly flat Mott lobes are in fact an artifact of our methods. 
There is two ways of introducing a more non-trivial $\mu_{\rm eff}$ dependence of the critical hopping going beyond those methods.
The first is going beyond a Lindblad equation \eqref{eq:singSiteME} to model the structured reservoir for the single-site problem (or also at lattice level (1)). In fact, that the bath spectral function is evaluated only at the excitation energies of the bare system  $E_{n+1} - E_{n} = nU$, is ultimately a result of the Born-Markov approximation. 
The second is to use approaches to the lattice problem that don't reduce to the single-site problem. In fact in the Lindblad equation for the lattice problem (1) the hopping $J$ enters (together with $\mu_{\rm eff}$) in the dissipator, as this is evaluated at the transition energies of the lattice Hamiltonian, rather than single-site one: this would introduce a non-trivial dependence of $J_c$ on $\mu_{\rm eff}$. 

While we expect some $\mu_{\rm eff}$-dependence beyond our approaches, we also expect that $\w_c$ will still depend strongly on the pump/loss ratio $r$ and be different from its ground-state value.

\subsection{Dynamical Mean-Field Theory}
\label{app:dmft}


The phase boundary condition can be obtained within DMFT using a similar procedure, as discussed in detail in Ref.~\cite{scarlatellaSchiro2021}. The key difference with respect to Gutzwiller/RPA is that the self-consistent symmetry breaking field takes contributions both from the coherent neighboring sites, as in Gutzwiller, as well as from the incoherent neighbors through the self-consistent DMFT bath. Assuming a Bethe lattice, the critical point equations \eqref{eq:neqCritFreq} \eqref{eq:neqCritHop} becomes in DMFT \cite{scarlatellaSchiro2021} 
\begin{align}
\label{eq:critFreqDmft}
\operatorname{Im} G^R\left(\omega_c, J_c\right)=0 \\ 
\label{eq:critHopDmft}
\frac{1}{J_c}+\operatorname{Re} G^R\left(\omega_c, J_c\right)+\frac{J_c}{z}\left[\operatorname{Re} G^R\left(\omega_c, J_c\right)\right]^2=0 
\end{align}
in terms of the local susceptibility $G^R(\w,J) = -i \int_{0}^\infty dt e^{i \w t } \aver{[ \hat{a}(t),\hat{a}^\da(0) ]}$.
We remark that the critical frequency $\w_c$ is still the zero of its imaginary part.
On the other hand, the local susceptibility here depends on the hopping and therefore the two equations are coupled, differently from the equations \eqref{eq:neqCritFreq},\eqref{eq:neqCritHop} where we could determine $\w_c$ from the first equation alone.

\section{Reservoir with Lorentzian correlations}
\label{app:lorentzianRes}
\begin{figure}
\centering
\includegraphics[width=1\linewidth]{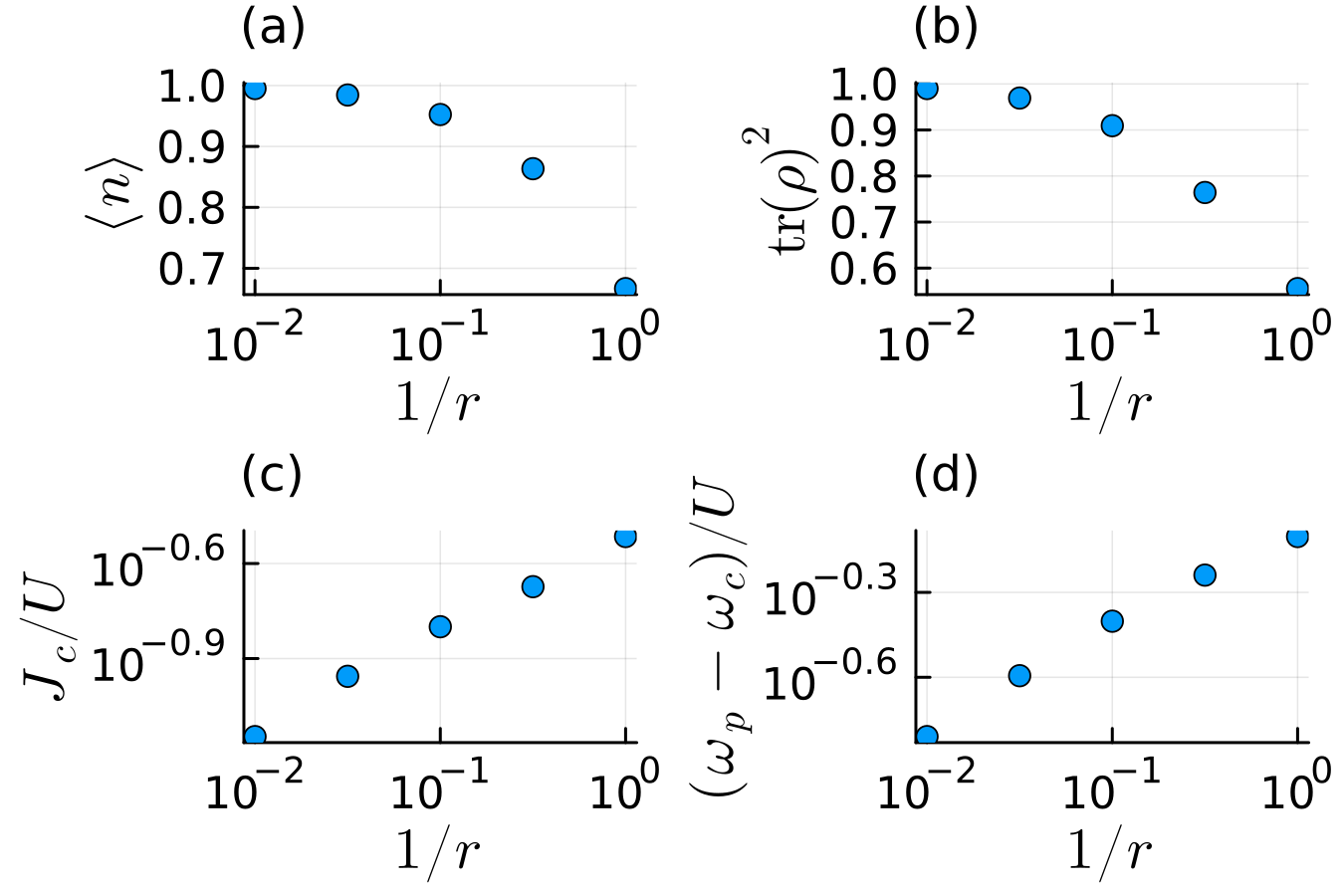}  
\caption{For a reservoir with Lorentzian spectral function \eqref{eq:lorCorr}, the driven-dissipative single-site steady-state average population $\aver{n}$, purity $\tr(\rho^2)$, critical hopping $J_c$ and frequency $\omega_c$ as a function of the pump-to-loss ratio $r$ (the doublon frequency is indicated by $\w_p$ here). 
The same qualitative behaviour found for the square-shaped reservoir spectral function of the main text is found. 
Parameters: $k/U = 10^{-6}$, $\gamma/U=10^{-3}$, $\omega_{\rm stab}/U = 1$.
}
\label{fig:jcWcLor}
\end{figure}


In the main text we considered a structured reservoir with a simple square-shaped correlation function  \eqref{eq:lesserBox}, acting as a chemical potential. 
Here we show that our conclusions do not depend on this specific choice, by considering instead a reservoir with a Lorentzian spectral function, which is qualitatively more similar to the experimental realization of Ref. \cite{maSchuster2019}. 
We assume the Lorentzian to be centred at $\w_{\rm res} = U$ corresponding to the transition from 0 to 1 photon in a Bose-Hubbard site with energy difference $E_1 - E_0 = U$ and to have a lifetime $\gamma$ (corresponding to the lifetime of reservoir excitations): 
\beq
\label{eq:lorCorr}
S^R(\w) =  r\kappa \frac{(\gamma/2)^2}{(\w-\w_{\rm res})^2 + (\gamma/2)^2}
\eeq

%

Fig. \ref{fig:jcWcLor} shows in panels (a-b) that upon increasing the pump/loss ratio the single-site steady-state approximates the Bose-Hubbard-site ground state with unit filling, and in panels (c-d) that we get a similar instability of the steady-state Mott phase with a similar behaviour of $\w_c$ and $J_c$ as shown in Fig. \ref{fig:phaseDiag} (d-e).

We also remark that the Markovian assumption, requiring that the timescale for the decay of the bath spectral function is shorter than the bath-induced system timescales, is strictly speaking not satisfied for the square-shaped correlation functions of the reservoir used in the main text \cite{lebreuillyCarusotto2017}. For the Lorentzian function used here instead the Markovian assumption is justified for our choice of parameters satisfying $\kappa, r \kappa \ll \gamma, \omega_{\rm res}$, showing that our conclusions are not affected.

\section{Redfield master equation}
\label{app:redfield}

\begin{figure}
\centering
\includegraphics[width=1\linewidth]{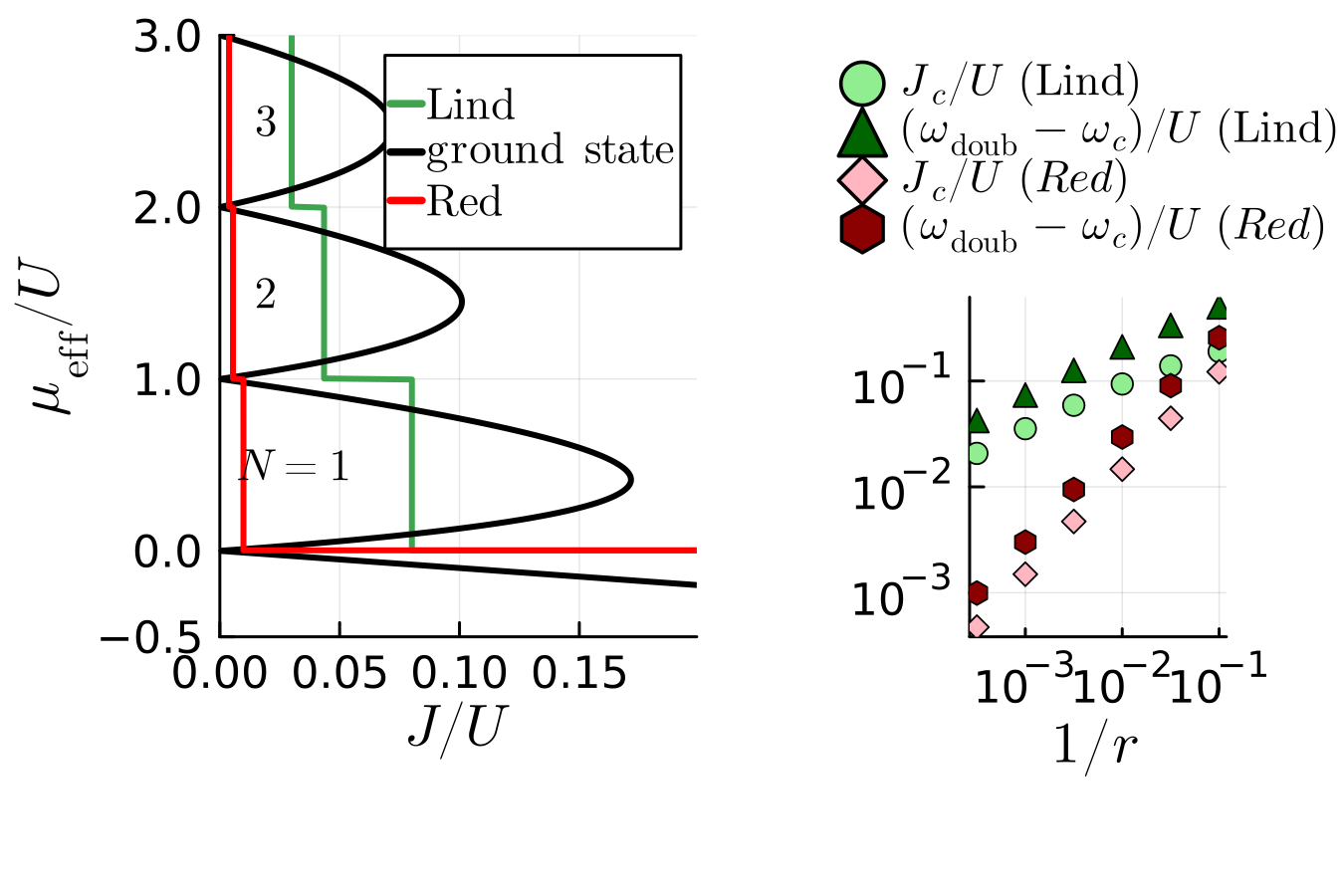}   
\caption{Steady-state Mott instability using the Lindblad \eqref{eq:mbME} versus Redfield \eqref{eq:redME} equation for the same parameters in Fig. \ref{fig:phaseDiag}. Left: steady-state phase diagram, where the ground-state Mott-superfluid phase diagram is plotted for comparison.
Right: critical hopping $J_c$ and frequency difference $\w_{\rm doub} - \w_c$ as a function of the inverse pump/loss ratio.
For the same parameters, the critical hopping is smaller in the Redfield case, as a feature of the doublon resonance that is crucial for the critical point appears at first order in perturbation theory in the dissipation strength in the Redfield case, while only at higher orders in the Lindblad case.
}
\label{fig:phaseDiag_red}
\end{figure}

In Ref. \cite{lebreuillyCarusotto2017} the same system considered in this paper is modelled using a Redfield master equation, rather than the Lindblad equation \eqref{eq:mbME}, which includes non-secular terms and Lamb-shift contributions to the Hamiltonian. 
In this appendix we show that our conclusions do not change considering such a Redfield equation. This reads
\beq
\partial_t \hat{\rho}=-i[\hat{H}, \hat{\rho}] + \kappa \sum_{j} \mathcal{D} [\hat{a}_{j}] \hat{\rho}  + \hat{\mathcal{D}}_{\rm stab}
\label{eq:redME}
\eeq
The structured reservoir dissipator in this case reads
\beq
\label{eq:mbDiss_red}
\hat{\mathcal{D}}_{\rm stab} = {r \kappa} \sum_i \lp a_i^\da \rho \tilde{a}_i  + \tilde{a}_i^\da \rho a_i - a_i \tilde{a}_i^\da \rho -\rho \tilde{a}_i a_i^\da \rp
\eeq
where the ``filtered'' operators $\tilde{a}_i = \sum_{m,n} S^R(\epsilon_n-\epsilon_m) \bra{\psi_m} a_i \ket{\psi_n} \ket{\psi_m} \bra{\psi_n}
$ are defined in terms of the eigenstates $\ket{\psi_n}$ and eigenvalues $\epsilon_n$ of the Bose-Hubbard Hamiltonian \eqref{eq:hamBh} and 
\beq
\label{eq:lesserBox_red}
S^R(\omega)  = \frac{1 }{2}  \theta(\mu_{\rm eff} - {\w}) \theta( \w +\w_0) +  \frac{i}{2\pi} \log \abs{\frac{\mu_{\rm eff} - \w}{\w_0+\w}} 
\eeq
whose imaginary part leads to a Lamb-shift term contributing to the Hamiltonian which is negligible in our results (for which the dissipation is small and staying away from the points in which this function is singular), but we kept it in the results of this appendix. 

Using such a Redfield equation \eqref{eq:redME}, the main results presented in the main text are recovered in Fig. \ref{fig:phaseDiag_red}, showing a similar phase diagram and a similar behaviour of $J_c$ and $\w_c$ at small $1/r$ as in Fig. \ref{fig:phaseDiag}.
The main difference is that the critical hopping is much smaller than in the Lindblad case and the exponent with which the critical hopping $J_c$ and the frequency difference $\w_{\rm doub}-\w_c$ decrease as a power law with $1/r$ is larger in absolute value compared to the Lindblad case.

In sections \ref{app:pertRed} and \ref{app:pertLind} we report perturbative calculations showing that this quantitative difference is due to the fact that a crucial contribution to the doublon resonance, that mainly determines the critical point, appears at first order in perturbation theory in the dissipation strength in the Redfield case, while for the Lindblad equation it only appears at higher orders in perturbation theory. 


We also remark that the Gutzwiller mean-field dynamics using the Redfield equation becomes unphysical in the limit-cycle phase, yielding negative probabilities, contrarily to the case of the Lindblad equation \eqref{eq:mbME} considered in the main text.

\section{The doublon resonance and the steady-state instability}
\label{app:doublonResonance}

\begin{figure}[b]
\centering
\includegraphics[width=1\linewidth]{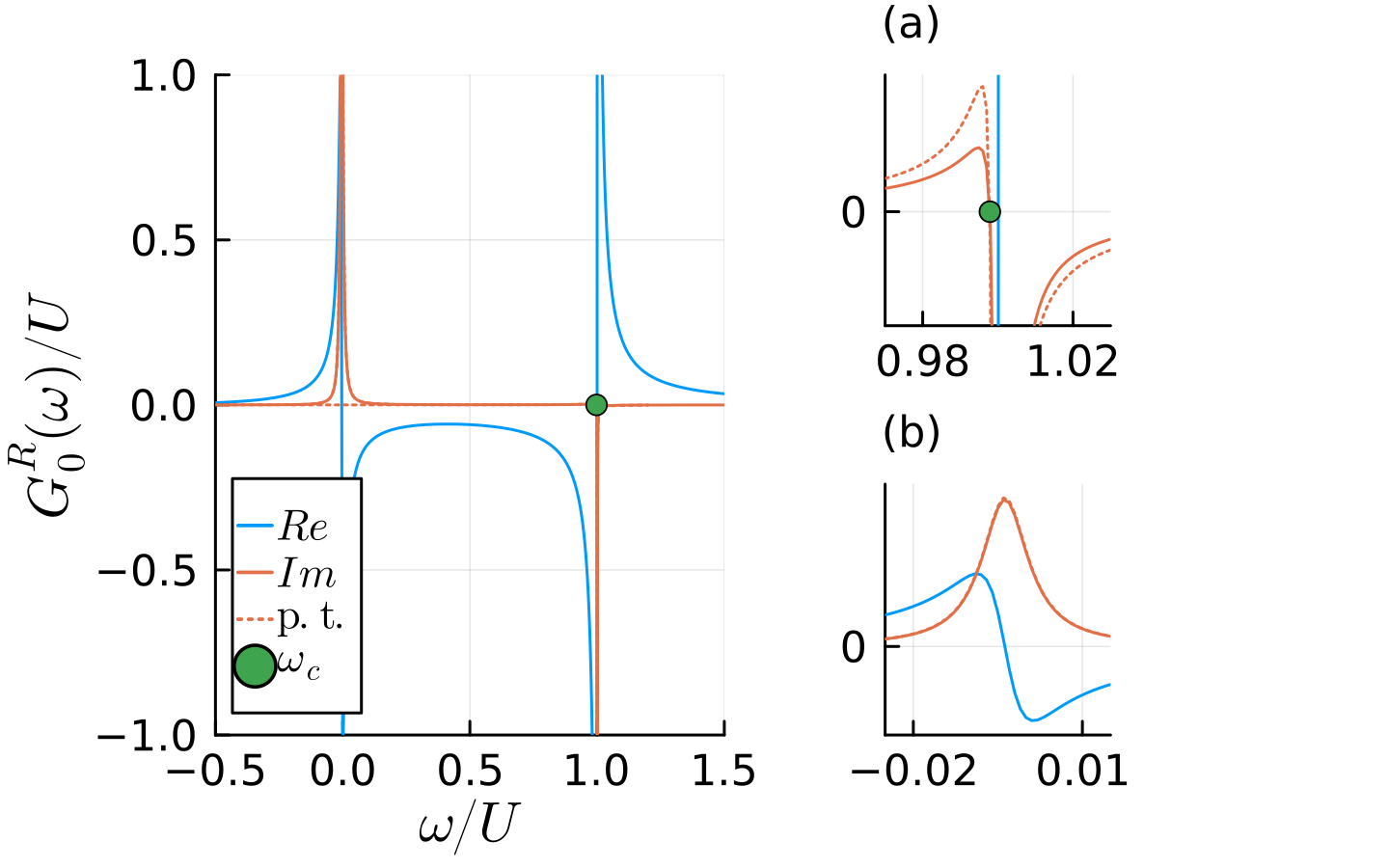}    
\caption{solid lines: single-site susceptibility for $r=10^3$ for the Redfield master equation \eqref{eq:redME} and for $\mu_{\rm eff}/U \in [0,1], \kappa/U = 10^{-5}$, obtained numerically. (a-b) zooms respectively around the right dobulon and left holon peak.
The doublon peak is strongly asymmetric due to an “anti-Lorentzian” contribution whose magnitude increases with the pump/loss ratio $r$.
The dotted lines are obtained using the first order perturbation theory in the dissipator, where the doublon peak is approximated by keeping only the corresponding term \eqref{eq:signPeak} in the Lehmann representation, showing that the zero of the imaginary part is correctly captured.
}
\label{fig:retGreen}
\end{figure}

The observation that a zero of the imaginary part of the single-site susceptibility $G_0^R(\w)$ forms close to its doublon peak (see the discussion of Fig. \ref{fig:dmft} (a)) allows to better understand the mathematical origin of the steady-state Mott instability. 
We find that for $\kappa , \kappa r\ll U,J$, where the peaks are well resolved, such a zero emerges due to an ``anti-Lorentzian'' contribution to $\im G_0^R(\w)$ \cite{scarlatellaSchiro2019a} making the doublon peak asymmetric, as Fig. \ref{fig:retGreen} shows plotting $ G_0^R(\w)$ in the case of the Redfield equation \eqref{eq:mbDiss_red} for which this behaviour is particularly pronounced (though for a Lindblad equation the same conclusions are true): in zoom (a) on the doublon resonance the imaginary part is clearly asymmetric, while the same asymmetry is not present in zoom (b) showing the holon resonance, whose imaginary part is almost even around the peak center. 

In the following, we discuss this anti-Lorentzian contribution and how it can lead to the steady-state instability, while in Sec. \ref{eq:mbDiss_red} we use first-order perturbation theory in the Redfield dissipator to capture this contribution analytically and in Sec. \ref{app:pertLind} we show that the same effect is not captured at first-order in perturbation theory in the Lindblad case (but appears at higher orders), explaining why the instability is somehow less pronounced in this case.

%

The retarded Green function (susceptibility) 
$G_0^R(t) = -i \tr \lbr  [ \hat{a}(t),\hat{a}^\da(0) ] \hat{\rho} \rbr \theta(t) $
of a problem described by a Markovian master equation $\pt \hat{\rho} = \hat{\mathcal{L}} \hat{\rho}$ can be decomposed in terms of the right $\hat{r}_\alpha$ and left $\hat{l}_\alpha$ eigenstates of the Lindblad superoperator $\hat{\mathcal{L}}$ and its eigenvalues $\lambda_\alpha$ in its Lehmann representation (see e.g. \cite{scarlatellaSchiro2019a}) 
\begin{equation}
\label{eq:retOpenSys}
G^R(\omega)=\sum_\alpha \frac{w_\alpha}{\omega+\operatorname{Im} \lambda_\alpha-\mathrm{iRe} \lambda_\alpha}
\end{equation}
with $w_\alpha=\operatorname{tr}\left(\hat{a} \hat{r}_\alpha\right) \operatorname{tr}\left(\hat{l}_\alpha^{\dagger}\left[\hat{a}^{\dagger}, \hat{\rho}\right]\right)$. 
It is important to notice that the weights $w_\alpha$ are in general complex, differently from the case of closed systems. 
The imaginary part takes the form
\beq
\label{eq:imGr}
\begin{split}
\im G^R(\w) = \sum_\alpha &\frac{\re w_\alpha \re \lambda_\alpha}{(\w + \im \lambda_\alpha)^2 + (\re\lambda_\alpha)^2} + \\ &\frac{\im w_\alpha}{\re \lambda_\alpha } \frac{\re \lambda_\alpha (\w + \im \lambda_\alpha)}{(\w + \im \lambda_\alpha)^2 + (\re\lambda_\alpha)^2}
\end{split}
\eeq
with both Lorentzian and ``anti-Lorentzian'' contributions corresponding to the first and second term. 
Note that in the limit of vanishing dissipation, where $\re \lambda_\alpha \rw 0$ and $\im w_\alpha \rw 0$, the amplitude of the anti-Lorentzian contribution $\im w_\alpha / \re \lambda_\alpha$ can still be finite. 

An important observation is that, in the regime of small dissipation $\kappa, \kappa r \ll U,J$ considered in the paper, the peaks stemming from different contributions in the sum are well separated in frequency, and therefore both the zero of the imaginary part and the critical hopping mostly originate solely from a single peak, the doublon peak (as we show in Fig. \ref{fig:retGreen}. This can be parametrized by 
\beq
f(\w) =b \frac{1 - i \gamma a}{\w + i \gamma} 
\eeq
with imaginary part 
\beq
\im f(\w) = -b \lp \frac{\gamma}{\w^2 +\gamma^2} + a \frac{\w \gamma}{\w^2 +\gamma^2} \rp
\eeq
The latter has a zero at $\w = -1/a$, that we identify with the deviation of the critical frequency from the doublon energy $ -1/a \approx \w_c - \w_{\rm doub} $. 
The critical hopping is then given by $J_c = -1/\re G^R(\w_c)$, that is by the real part of the $f(\w)$
\beq
\re f(\w) = b \lp \frac{\w }{\w^2 +\gamma^2} - \frac{ \gamma^2 a }{\w^2 +\gamma^2} \rp
\eeq
yielding $J_c \propto -1/\re f(-1/a) = 1/(a b) $. 
Therefore we get the proportionality $\w_{\rm doub} - \w_c \propto J_c$, explaining why those quantities vanish simultaneously for large $r$, as shown in Fig. \ref{fig:phaseDiag}.(c-d). 
Another way to understand this proportionality is that the real and imaginary part of the susceptibility are related by the Kramers-Kronig relations \cite{colemanColeman2015}. 


\onecolumngrid
\subsection{Perturbation theory for the Redfield equation}
\label{app:pertRed}

We now show, using perturbation theory, that a strong anti-Lorentzian contribution proportional to $r$ indeed arises in correspondence of the doublon peak, giving rise to a zero of $\im G_0^R(\w)$ setting the critical frequency \eqref{eq:neqCritFreq}, and not in correspondence of other resonances of the Green function. 
%
%
%
In order to capture this feature, we evaluate the Lehmann representation \eqref{eq:retOpenSys} 
using first-order perturbation theory in the dissipators to approximate the eigenstates and eigenvalues of the Liouvillian. We define $\mathcal{L} = -i \lsq H, \bullet \rsq + \mathcal{D}$ and we perturb in the second term, as for example in \cite{scarlatellaSchiro2019a}. 
The unperturbed eigenstates and eigenvalues are $\lambda_{n,m}^{(0)} = -i (E_n - E_m) $, $r^{(0)}_{n,m} = l^{(0)}_{n,m}= \ket{n}\bra{m}$. 
First-order perturbation theory gives the following corrections to eigenvalues and eigenstates

\begin{align}
\lambda_\alpha^{(1)}&=\operatorname{tr}[ ({l}_\alpha^{(0)} )^{\dagger} {\mathcal{D}} ( {r}_\alpha^{(0)} ) ] & 
{r}_\alpha^{(1)}&=\sum_{\beta \neq \alpha} \frac{\operatorname{tr}[({r}_\beta^{(0)})^{\dagger} {\mathcal{D}}({r}_\alpha^{(0)})]}{\lambda_\alpha^{(0)}-\lambda_\beta^{(0)}} {r}_\beta^{(0)} &
{l}_\alpha^{(1)}&=\sum_{\beta \neq \alpha} \frac{\operatorname{tr}[({l}_\beta^{(0)})^{\dagger} {\mathcal{D}}^{\dagger}({l}_\alpha^{(0)})]}{\lambda_\alpha^{(0) *}-\lambda_\beta^{(0) *}} {l}_\beta^{(0)}
\end{align}

with $\alpha = (n,m)$. 

The term $\tr ( a r_{n,m} )$ in the Green function weights $w_{n,m}$ selects only the eigenstates/values with $m+1=n$, thus we compute only those eigenvalues/states, obtaining
\small
\begin{align} 
r_{n+1,n} &\approx \ket{n+1} \bra{n} + i \frac{\kappa}{U} \sqrt{(n+1)n} \ket{n} \bra{n-1} - i \frac{r\kappa}{U} \sqrt{(n+2)(n+1)} \lsq S^R_{+-}(E_{n+1} -E_{n}) + S^R_{+-}(E_{n+2} -E_{n+1})^* \rsq \ket{n+2}\bra{n+1} \label{eq:pertVecs_1}\\ 
l_{n+1,n} &\approx \ket{n+1} \bra{n} + i \frac{\kappa}{U} \sqrt{(n+2)(n+1)} \ket{n+2}\bra{n+1} - i \frac{r \kappa}{U} \sqrt{(n+1)n} \lsq S^R_{+-}(E_{n+1} -E_{n})  + S^R_{+-}(E_{n} -E_{n-1})^* \rsq \ket{n} \bra{n-1} \label{eq:pertVecs_2}\\ 
\lambda_{n+1,n} &\approx -i (E_{n+1} -E_n) - \frac{\kappa}{2} (2n+1) - {r\kappa} \lsq S^R_{+-}(E_{n+2} -E_{n+1})^* (n+2) + S^R_{+-}(E_{n+1} -E_{n}) (n+1)  \rsq \label{eq:pertVals} 
\end{align}
\normalsize

%
To determine the amplitude of the anti-Lorentzian contribution given by $\im w_{n+1,n}/ \re \lambda_{n+1,n} $, we first compute 
\footnotesize
\beq
\es{
\im w_{n+1,n} &= \sqrt{n+1} \lsq \frac{\kappa}{U}(n+2)\sqrt{n+1} \lp p_{n+1} -p_{n+2} \rp -\frac{r\kappa}{U}  \sqrt{(n+1)n} \lp \re S^R(E_{n+1}-E_n)^* + \re S^R(E_{n}-E_{n-1}) \rp \lp p_{n-1} \sqrt{n} -p_n\sqrt{n+1}\rp \rsq \\ 
&+ (p_n - p_{n+1}) \sqrt{n+1} \lsq \frac{\kappa}{U} n \sqrt{n+1} -\frac{r\kappa}{U} \sqrt{n+1}(n+2) \lp \re S^R(E_{n+1}-E_n) + \re S^R(E_{n+2}-E_{n+1})^* \rp \rsq
}
\eeq
\normalsize 

Then we check for which values of $n$ (i.e. in correspondence of which resonance of the Green function) $\im w_{n+1,n}/ \re \lambda_{n+1,n} $ is of order $r$.
We use that the spectral function of the reservoir $S^R(E_{n}-E_{n-1}) = \frac{1}{2} \theta(\sigma - \abs{E_{n}-E_{n-1}} )$ \eqref{eq:lesserBox_red} (we neglect its imaginary part here) vanishes for $n>N$.
In $\im w_{n+1,n}$ all the terms proportional to $r$ vanish for $n>N$, while $\re \lambda_{n+1,n}$ has no terms proportional to $r$ for $n>N-1$. Then for $n=N$ and only in this case the ratio $\im w_{N+1,N}/ \re \lambda_{N+1,N} $ is proportional to $r$. 
The contribution for $n=N$ corresponds to the doublon resonance, that therefore acquires an anti-Lorentzian contribution proportional to $r$ that eventually leads to a critical point \eqref{eq:neqCritFreq}\eqref{eq:neqCritHop}.  

%
%


Eventually, we can approximate the Green function peak for $n=N$ with the expression
\beq
\label{eq:signPeak}
G_{N+1,N}^R(\w) = \frac{\lp \sqrt{N+1} + \frac{i\kappa}{U} \sqrt{(N+1)}N \rp 
\lp p_N  \sqrt{N+1} - i \frac{r\kappa}{U} p_{N} (N+1)\sqrt{N}  \rp}{\w  - (E_{N+1}-E_N) +\frac{i\kappa}{2}(2N+1) }
\eeq
where we also used that $p_{n\neq N} \approx 0$ for $r \gg 1$.
In Fig. \ref{fig:retGreen} we plot this single-peak approximation as a dashed line for the doublon peak (panel (a)) and show that it correctly captures the anti-Lorentzian, thus the critical frequency $\w_c$ and hopping $J_c$. The dashed line in panel (b) instead approximate the hole-like resonance using the full Green function in perturbation theory. 

We remark that the precise square shape of the reservoir spectral function is not important, as long as this function strongly suppress transitions above a certain energy. Indeed the same behaviour is observed in this Supplemental Material for a reservoir with Lorentzian spectral function.


\subsection{Perturbation theory for the Lindblad equation}
\label{app:pertLind}

The perturbation theory using the Lindblad equation \eqref{eq:mbME} instead of Redfield \eqref{eq:redME} is very similar, and one only needs to discard the non-secular terms that are present in the latter case but not in the former.
Discarding those terms yields the same first-order corrections for the eigenvalues of the Liouvillian  \eqref{eq:pertVals} as in Redfield, while all the eigenstates corrections in \eqref{eq:pertVecs_1},\eqref{eq:pertVecs_2} coming from the pump term (proportional to $r \kappa$) correspond to non-secular terms and thus vanish. Therefore the spectral function weights $w_{n+1,n}$ (depending on the eigenstates) also do not depend on $r$ for all $n$ and, eventually, the amplitude of the anti-Lorentzian contributions given by the ratio $\im w_{n+1,n}/ \re \lambda_{n+1,n} $ is never proportional to $r$.

Note though that the same behaviour of the critical frequency $\w_c$ approaching the doublon energy $ \w_{\rm doub}$ increasing the pump/loss ratio $r$ is observed for the Lindblad case (Fig. \ref{fig:phaseDiag}(e)), therefore a similar anti-Lorentzian contribution to the doublon resonance proportional to $r$ is expected to arise from higher-order terms in perturbation theory.

This difference between the Redfield and Lindblad perturbation theories explains why the dissipative-Mott instability is more pronounced in the former case, with a smaller critical hopping, while both equations give qualitatively the same predictions. 

\twocolumngrid


%

\end{document}